\documentclass[12pt,twoside]{article}
\usepackage{fleqn,espcrc1}

\usepackage{graphicx}
\usepackage[figuresright]{rotating}

\title{Dyson--Schwinger studies of meson masses and decay constants}

\author{P. Maris\address{Dept. of Physics, 
        Kent State University, Kent OH 44242}%
        }

\begin{document}

% typeset front matter
\maketitle

\begin{abstract}
The masses and decay constants of the light mesons are studied within a
ladder-rainbow truncation of the set of Dyson--Schwinger equations using
a model 2-point gluon function.  The one phenomenological parameter and
two current quark masses are fitted to reproduce $f_\pi$, $m_\pi$ and
$m_K$.  Our results for $f_K$, and for the vector mesons $\rho$, $\phi$,
and $K^\star$ are in good agreement with the experimental values.
\end{abstract}

\section{DYSON--SCHWINGER EQUATIONS}

The Dyson--Schwinger equations [DSEs] form an excellent tool to study
nonperturbative aspects of quark propagators and the formation of bound
states~\cite{dserev}.  The approach is consistent with quark and gluon
confinement~\cite{confinement} and generates dynamical chiral symmetry
breaking.  It is straightforward to implement the correct one-loop
renormalization group behavior of QCD~\cite{MR97}, and obtain agreement
with perturbation theory in the perturbative region.  Phenomenological
models based on the DSEs~\cite{PCT97} have recently allowed significant
progress not only for light mesons, but also for heavy
mesons~\cite{IKR99}.

The DSE for the renormalized dressed-quark propagator in Euclidean
formulation is
\begin{eqnarray}
\label{gendse}
 S(p)^{-1} & = & Z_2\,i\gamma\cdot p + Z_4\,m_q(\mu)
        + Z_1 \int^\Lambda\!\frac{d^4q}{(2\pi)^4} g^2 D_{\mu\nu}(p-q) 
        \frac{\lambda^a}{2}\gamma_\mu S(q)\Gamma^a_\nu(q,p) \,,
\end{eqnarray}
where $D_{\mu\nu}(k)$ is the renormalized dressed-gluon propagator,
$\Gamma^a_\nu(q;p)$ the renormalized dressed-quark-gluon vertex, and
$\int^\Lambda$ denotes a translationally-invariant regularization of the
integral with $\Lambda$ the regularization mass-scale.  The general
solution of Eq.~(\ref{gendse}) has the form $ S(p)^{-1} = i \gamma\cdot
p A(p^2) + B(p^2)$, renormalized according to
\begin{equation}
\label{renormS}
S(p)^{-1}\bigg|_{p^2=\mu^2} = i\gamma\cdot p + m_q(\mu)\,,
\end{equation}
with $m_q(\mu)$ the current quark mass at the renormalization scale
$\mu$.  The renormalization constants $Z_1$, $Z_2$ and $Z_4$ generally
depend on $\mu$ and $\Lambda$, but not on the quark flavor.

The renormalized, homogeneous Bethe--Salpeter equation [BSE] for a bound
state of a quark of flavor $a$ and an antiquark of flavor $b$ having
total momentum $P$ is given by
\begin{equation}
\label{hombse}
 \Gamma^{ab}_M(p;P) =  \int^\Lambda\!\frac{d^4q}{(2\pi)^4} K(p,q;P) 
        S^a(q+\eta P) \Gamma^{ab}_M(q;P) S^b(q-\bar\eta P)\;, 
\end{equation}
where \mbox{$\eta + \bar\eta = 1$} describes momentum sharing,
$\Gamma^{ab}_M(p;P)$ is the Bethe--Salpeter amplitude [BSA], and $M$
specifies the meson type: pseudoscalar, vector, axial-vector, or scalar.
The kernel $K$ operates in the direct product space of color and Dirac
spin for the quark and antiquark and is the renormalized, amputated
$\bar q q$ scattering kernel that is irreducible with respect to a pair
of $\bar q q$ lines.  This equation defines an eigenvalue problem with
physical solutions at the mass-shell points \mbox{$P^2=-m^2$} with $m$
being the bound state mass.

We use a ladder truncation for the BSE 
\begin{equation}
\label{ourBSEansatz}
        K^{rs}_{tu}(p,q;P) \to
        -{\cal G}((p-q)^2)\, D_{\mu\nu}^{\rm free}(p-q)
        \left(\frac{\lambda^a}{2}\gamma_\mu\right)^{ru} \otimes
        \left(\frac{\lambda^a}{2}\gamma_\nu\right)^{ts} \,,
\end{equation}
where $D_{\mu\nu}^{\rm free}(k)$ is the bare gluon propagator in Landau
gauge.  Together with a rainbow truncation
\mbox{$\Gamma^a_\nu(q,p) \rightarrow \gamma_\nu \lambda^a/2$} for the quark 
DSE, Eq.~(\ref{gendse}), this preserves the axial-vector Ward--Takahashi
identity~\cite{BRvS96}, which ensures that in the chiral limit the
ground state pseudoscalar mesons are massless even though the quark mass
functions are strongly enhanced in the infrared.  Thus pions are
understood as both the Goldstone bosons associated with dynamical chiral
symmetry breaking and as $\bar q q$ bound states~\cite{MR97,MRT98}.

For the ``effective coupling'' ${\cal G}(k^2)$ we employ the
Ansatz~\cite{MT99}
\begin{equation}
\label{gvk2}
\frac{{\cal G}(k^2)}{k^2} =
               \frac{4\pi^2}{\omega^6} D k^2 {\rm e}^{-k^2/\omega^2}
        + 4\pi\,\frac{ \gamma_m \pi}{\frac{1}{2}
        \ln\left[\tau + \left(1 + k^2/\Lambda_{\rm QCD}^2\right)^2\right]}
        {\cal F}(k^2) \,,
\end{equation}
with ${\cal F}(k^2)= [1 - \exp(-k^2/[4 m_t^2])]/k^2$, $\tau={\rm
e}^2-1$, and \mbox{$\gamma_m=12/(33-2N_f)$}.  This Ansatz preserves the
one-loop renormalization group behavior of QCD; in particular, one
obtains the correct one-loop QCD anomalous dimension of the quark mass
function $M(p^2):= B(p^2)/A(p^2)$ for both the chiral limit ($m(\mu)=0$)
and the explicit chirally broken case ($m(\mu) \neq 0$).  The first term
of Eq.~(\ref{gvk2}) implements a strong infrared enhancement in the
region \mbox{$k^2<1$}~GeV$^2$ which is phenomenologically required for
sufficient dynamical chiral symmetry breaking and for
confinement~\cite{HMR98}. We use \mbox{$m_t=0.5$~GeV},
\mbox{$N_f=4$}, \mbox{$\Lambda_{\rm QCD}^{N_f=4}= 0.234\,{\rm GeV}$},
\mbox{$\omega = 0.4$~GeV}, and a renormalization point $\mu=19\,$GeV.

\section{MESON BETHE--SALPETER AMPLITUDES}

The BSA $\Gamma_{PS}(q;P)$ of a pseudoscalar meson has the general form
\begin{equation}
\label{genpion}
 \Gamma_{PS}(q;P)  =  \gamma_5 \left( i E + 
        \gamma\cdot P \, F + \gamma\cdot q \,\hat{G}
        + \sigma_{\mu\nu}\,q_\mu P_\nu \,H \right)\,,
\end{equation}
with the invariant amplitudes $E$, $F$, $\hat{G}$ and $H$ being Lorentz
scalar functions of $q^2$ and $q\cdot P$; $P^2= -m_{PS}^2$ is the fixed,
on-shell, meson momentum.  For unflavored mesons that are charge
conjugation eigenstates, such as the $\pi$, there is an additional
constraint on the BSA to obtain a specified $C$-parity\footnote{We do
not discriminate between up and down quarks, nor incorporate
electromagnetic corrections; therefore the BSA for $\pi^\pm$ is equal to
that for $\pi^0$.}.  Of the four covariants given in
Eq.~(\ref{genpion}), $\hat{G}$ is odd under $C$, the others are even
under $C$.  The only remaining quantity that can produce a desired
uniform $C$-parity is $q \cdot P$ which is odd under $C$.  Thus a pion
will have an amplitude $\hat{G}$ that is odd in $q\cdot P$ while the
other amplitudes are even in $q\cdot P$.  Flavored mesons like kaons,
will have amplitudes that are neither even nor odd in $q\cdot P$.

Massive vector mesons are transverse, and the general form of the BSA
$\Gamma_\mu^V(q;P)$ can be expressed in terms of eight independent
Lorentz covariants $T^i_\mu(q;P)$ with coefficients $F_i(q^2, q \cdot P;
P^2)$.  The choice for the covariants $T^i_\mu(q;P)$ to be used as a
basis is constrained by the required properties under Lorentz and parity
transformations, but is not unique; a convenient orthogonal basis is
given in Ref.~\cite{MT99}.  For the other mesons, scalar and
axial-vector, covariants can be defined similarly.  However, we expect
the ladder truncation used here to be reliable for pseudoscalar and
vector, but not for scalar mesons~\cite{BRvS96,R97ladsc}.

The electromagnetic decay mediated by a photon (e.g. $\rho^0$, $\omega$,
$\phi$), and the leptonic decay mediated by a W-boson (e.g. $\rho^\pm$,
$K^{\star\pm}$), are described by the vector decay
constant~\cite{IKR99,MT99,GL82}
\begin{eqnarray}
   f_V\, m_V\, \epsilon_\mu^{(\lambda)}(P) &=& 
        \langle 0|\bar q^b \gamma_\mu q^a| V^{ab}(P,\lambda)\rangle \,,
\end{eqnarray}
where $\epsilon_\mu^{(\lambda)}$ is the polarization vector of the
vector meson.  This is analogous to
\begin{eqnarray}
   f_{PS}\, P_\mu\, &=& 
        \langle 0|\bar q^b \gamma_\mu \gamma_5 q^a| PS^{ab}(P)\rangle \,,
\nonumber\\
\label{psdecayconst}
         &=& {Z_2\,N_c}
        \int^\Lambda\!\!\frac{d^4q}{(2\pi)^4}\,
        {\rm Tr}_{D}\left[\gamma_\mu\gamma_5 S^a(q+\eta P) 
        \Gamma^{ab}_{PS}(q;P) S^b(q-\bar{\eta}P)\right]\,,
\end{eqnarray}
for the pseudoscalar decay constant corresponding to $f_\pi = 131$~MeV
under the normalization convention used here.  The vector decay constant
can be expressed as a loop integral
\begin{eqnarray}
\label{vecdecayconst}
   f_V\, m_V &=& \frac{Z_2\,N_c}{3} 
        \int^\Lambda\!\!\frac{d^4q}{(2\pi)^4}\,
        {\rm Tr}_{D}\left[\gamma_\mu S^a(q+\eta P) 
        \Gamma^{ab}_\mu(q;P) S^b(q-\bar{\eta}P)\right]\,.
\end{eqnarray}
Both Eq.~(\ref{psdecayconst}) and(\ref{vecdecayconst}) are exact if the
dressed quark propagators and the BSA are exact~\cite{IKR99,MRT98}.

\section{RESULTS FOR PSEUDOSCALAR AND VECTOR MESONS}

After insertion of the general form of $\Gamma_{PS}(q;P)$ or
$\Gamma_{\mu}^{V}(q;P)$ in terms of a covariant basis, the homogeneous
BSE Eq.~(\ref{hombse}), can be reduced to a set of coupled integral
equations for the scalar functions $F_i^{ab}(q^2,q\cdot P;P^2)$
\begin{eqnarray}
\label{ampleqn}
 F^{ab}_i(p^2,p\cdot P;P^2) &=& \int^\Lambda\!\!\frac{d^4q}{(2\pi)^4}\,
        {\cal K}^{ab}_{i,j}(p^2,p\cdot P;q^2,q\cdot P; p\cdot q;P^2)
        F^{ab}_j(q^2, q\cdot P;P^2) \,.
\end{eqnarray}
This equation is solved numerically without further approximations as an
eigenvalue problem to give the mesons masses; the decay constants follow
from Eqs.~(\ref{psdecayconst}) and (\ref{vecdecayconst}).  

The parameter $D$ and the current quark mass $m_{u/d}(\mu)$ are fixed by
fitting $m_\pi$ and $f_\pi$.  Next, the strange quark mass $m_{s}(\mu)$
is determined by a fit to $m_K$.  With \mbox{$D = 0.93\;{\rm GeV}^2$}
and quark masses \mbox{$m_{u/d}(\mu) = 5.5\;{\rm MeV}$}, and
\mbox{$m_s(\mu) = 130\;{\rm MeV}$} scaled to \mbox{$\mu = 1\;{\rm GeV}$}, 
we obtain a good description of the $\pi$, $\rho$, $K$, $K^\star$ and
$\phi$ masses and decay constants, see Table~\ref{table:1}.  Our results
are not very sensitive to the details of the gluon Ansatz~\cite{MT99},
similar results are found with $\omega = 0.3$~GeV and $\omega =
0.5$~GeV, as long as $D$ and the quark masses are refitted.  Also
$r_\pi$ is in excellent agreement with the experimental
value~\cite{MT99panic}.

\begin{table}[htb]
\caption{The pseudoscalar and vector meson masses $m_M$ 
and decay constants $f_M$ in GeV.}
\label{table:1}
\newcommand{\m}{\hphantom{$1$}}
\newcommand{\pfi}{\hphantom{$5$}}
\newcommand{\pse}{\hphantom{$7$}}
\newcommand{\cc}[1]{\multicolumn{1}{c}{#1}}
\renewcommand{\tabcolsep}{2pc} % enlarge column spacing
\renewcommand{\arraystretch}{1.2} % enlarge line spacing
\begin{tabular}{lcccc}
\hline
                & \cc{$m$ (calc.)}& \cc{$f$ (calc.)}
                & \cc{$m$ (exp.)} & \cc{$f$ (exp.)} \\
\hline
pion            & 0.138 & 0.131 & 0.1385   & 0.1307 \\
kaon            & 0.497 & 0.155 & 0.496\pfi& 0.160\pse\\
%$(\bar s s)_{PS}$& ???  &  ???  & \m---    & ---  \\ 
\hline
$\rho$          & 0.742 & 0.207 & 0.770\pfi& 0.216\pfi\\
$K^\star$       & 0.936 & 0.241 & 0.892\pfi& 0.225\pfi\\
$\phi$          & 1.072 & 0.259 & 1.020\pfi& 0.237\pfi\\
\hline
\end{tabular}
\end{table}

For small current quark masses, the pseudoscalar meson mass grows like
$m_{PS}^2\propto m_q(\mu)$, in agreement with the
Gell-Mann--Oakes--Renner relation.  For larger quark masses, $m_q(\mu) >
3\,m_s(\mu)$, the meson mass grows linearly with the current quark
mass~\cite{MR98heavy}, as expected for heavy mesons.  The decay
constants also increase with increasing quark masses, as can be seen
from Table~\ref{table:1}.  However, for heavy mesons it follows from the
general form of the BSA, together with the normalization condition, that
both the vector and the pseudoscalar decay constant decrease with the
quark mass as $f_M \propto 1/\sqrt{m_M}$~\cite{IKR99}.

For mesons that are charge conjugation eigenstates, the dependence of
the amplitudes $F_i(q^2,q\cdot P;P^2)$ on $q\cdot P$ is minimal.  For
flavored mesons there is a much stronger angular dependence.  Our
results for physical observables, such as the mass and decay constant,
are {\em independent} of the momentum sharing and of the regularization
scale $\Lambda$, as long as all relevant covariants and the full angular
dependence are included in the calculation.

\subsection*{Acknowledgments}

I would like to thank the organizers of PANIC99 and M.~Birse for
inviting me to give a talk at the parallel session on hadron
spectroscopy.  Most of this work was done in collaboration with
C.D.~Roberts and P.C.~Tandy.  It was funded in part by the National
Science Foundation under grant no PHY97-22429 and benefited from the
resources of the National Energy Research Scientific Computing Center.

\end{document}